\documentclass[prb,twocolumn,showpacs,superscriptaddress,preprintnumbers,amsmath,amssymb]{revtex4-1}
\usepackage{graphicx} \usepackage{dcolumn} \usepackage{bm} \usepackage{amsmath} \usepackage{color}
\usepackage{hyperref} \usepackage[utf8]{inputenc} \usepackage[T1]{fontenc}

\begin{document}


\title{Dynamical lattice instability versus spin liquid state in a frustrated spin chain system}

\author{V. Gnezdilov} \affiliation{Institute for Condensed Matter
Physics, Technical University of Braunschweig, D-38106 Braunschweig, Germany}
\affiliation{B.I. Verkin Inst. for Low Temperature Physics and
Engineering, NASU, 61103 Kharkov, Ukraine}

\author{P. Lemmens} \affiliation{Institute for Condensed Matter
Physics, Technical University of Braunschweig, D-38106 Braunschweig, Germany}

\author{Yu. G. Pashkevich} \affiliation{A.A. Galkin Donetsk Phystech, NASU, 83114 Donetsk, Ukraine}

\author{D. Wulferding} \affiliation{Institute for Condensed Matter
Physics, Technical University of Braunschweig, D-38106 Braunschweig, Germany}

\author{I.V. Morozov} \affiliation{M.V. Lomonosov Moscow State University, Moscow 119991, Russia}

\author{O.S. Volkova} \affiliation{M.V. Lomonosov Moscow State University, Moscow 119991, Russia}

\author{A. Vasiliev} \affiliation{M.V. Lomonosov Moscow State University, Moscow 119991, Russia}

\date{\today}

\begin{abstract} The low-dimensional $s=1/2$ compound (NO)[Cu(NO$_3$)$_3$] has recently been suggested to follow the Nersesyan-Tsvelik model of coupled spin chains. Such a system shows unbound spinon excitations and a resonating valence bond ground state due spin frustration. Our Raman scattering study of (NO)[Cu(NO$_3$)$_3$] demonstrates phonon anomalies as well as the suppression of a broad magnetic scattering continuum for temperatures below a characteristic temperature, $T<T^*\approx$ 100~K. We interpret these effects as evidence for a dynamical interplay of spin and lattice degrees of freedom that might lead to a further transition into a dimerized or structurally distorted phase at lower temperatures. \end{abstract}




\maketitle


\section{Introduction}
Quantum spin systems with competing interactions and reduced dimensionality of the spin exchange show exotic physics.~\cite{book-frust} In particular, there is the opportunity to observe unusual spin liquid ground states, exotic fractional excitations and to test various theoretical concepts. This has triggered intensive research activities. However, the realization of such states is challenging, as more conventional, long range ordered or dimerized (spin-Peierls) phases exist as competing instabilities. The needed fine tuning and tailoring of the crystallographic and electronic properties requires a considerable knowledge of materials aspects as well as deep theoretical understanding. So far, most of the studied systems can be classified as quasi 1-dimensional (1D) quantum spin chains, coupled chains, 2-dimensional (2D) triangular and kagome lattices.

The recently synthesized low dimensional compound nitrosonium nitratocuprate (NO)[Cu(NO$_3$)$_3$] has been discussed in the framework of a frustrated, quasi-2D systems without long range magnetic ordering.~\cite{volkova-10} The crystal structure of (NO)[Cu(NO$_3$)$_3$] consists of layers of CuO$_6$ octahedra that are coupled along the crystallographic $b$ direction by triangular (NO$_3$)$^-$ (nitrate) groups, see the inset in Fig.~\ref{overview+structure}. These chains are weakly coupled via double groups of singly connected nitrates. In addition (NO)$^+$ (nitrosonium) cations are located between the chains. Interestingly, these three structural subunits not only lead to exotic magnetic properties but also to phonon modes with three frequency ranges and a characteristic low energy onset, similar to fingerprints, as shown further below.

The magnetic susceptibility, specific heat, and electron spin resonance (ESR) measurements in (NO)[Cu(NO$_3$)$_3$] evidence pronounced spin fluctuations without any hint for long-range magnetic order down to 1.8 K.~\cite{volkova-10} A broad maximum in the magnetic susceptibility $\chi(T)$ is observed around $T^* \sim 100$ K, characterizing the energy scale of the spin exchange. There is no evidence for an excitation gap of the triplet states (spin gap) that would lead to a sharp suppression of $\chi(T)$ at lowest temperatures. However, $\chi(T)$ with $T<$$T^*$ is smaller than expected for an isolated $s=1/2$ Heisenberg chain.~\cite{volkova-10,janson-10}

From a detailed analysis of $\chi(T)$ it has been deduced that (NO)[Cu(NO$_3$)$_3$] is a realization of the 2D confederate flag or Nersesyan-Tsvelik model.~\cite{nersesyan-03,tsvelik04} This model consists of spin chains with additional direct and diagonal coupling between neighboring chains. The diagonal couplings are related to spin frustration as not all exchange paths can be satisfied simultaneously.~\cite{book-frust} The particular exchange geometry is known to lead either to a resonating valence bond (RVB)~\cite{book-frust} or a valence bond crystal (VBC) ground state with a macroscopic ground state degeneracy. These states support free, propagating spinons and more exotic topological excitations. The assignment of the compound to the Nersesyan-Tsvelik model is based on a critical ratio of three exchange coupling constants: the leading magnetic exchange coupling along the chains, $J$=170 K, the interchain coupling, $J'$, with $-8.5$ K $< J' <15.3$ K, and the exchange interaction $J_2$ along the diagonals. The latter is proposed to be exactly $1/2 \cdot J$, which leads to strong spin frustration, suppresses long-range order (LRO), and allows quantum fluctuating and entangled ground states. However, in the proximity to the critical coupling also dimerized phases exist.~\cite{tsvelik04}


In a recent density-functional-theory (DFT) band structure calculation doubts about the Nersesyan-Tsvelik model for (NO)[Cu(NO$_3$)$_3$] were raised.~\cite{janson-10} Considering a different geometry of superexchange pathways, a similar, leading antiferromagnetic exchange coupling was found with $J = 200$ K. However, much weaker couplings with $J' = 2$ K link the chains into layers. Therefore the compound would be better described as a uniform spin-chain with weak and nonfrustrated interchain coupling. Irrespective to the DFT the present thermodynamic experiments do not allow a clear consensus which model leads to a better description.

For a deeper understanding of the title compound it is of pivotal importance to probe both lattice and magnetic degrees of freedom using a spectroscopic experimental techniques.~\cite{lemmens-rev,lemmens-choi-enc} In addition, a previous Raman scattering study of (NO)[Cu(NO$_3$)$_3$] reported only four of the expected 45 phonon modes which further motivates a refined study.~\cite{givan-89} With our Raman experiment we support a frustrated spin model with a coupling constant of $J_b$$\approx$150~K and give further evidence for strong phonon anharmonicity in (NO)[Cu(NO$_3$)$_3$] possibly leading to a dynamic dimerization.

\section{Experimental details}

Single crystals of nitrosonium nitratocuprate (NO)[Cu(NO$_3$)$_3$] were obtained by means of a wet chemistry route.~\cite{znamenkov-04} The crystals are unstable under ambient conditions and were therefore kept in vacuum to prevent degradation of the freshly cleaved crystal surface during the investigation. Raman scattering experiments were performed in quasi-backscattering geometry using a $\lambda = 532$ nm solid state laser. The laser power was set to 3 mW with a spot diameter of approximately 100 $\mu$m to avoid heating effects. All measurements were carried out in an evacuated closed cycle cryostat in the temperature range from 10 K to 295 K. The spectra were collected via a triple spectrometer (Dilor-XY-500) by a liquid nitrogen cooled CCD (Horiba Jobin-Yvon, Spectrum One CCD-3000V).

\section{Experimental Results}

At temperatures above 160 K, the crystal structure of (NO)[Cu(NO$_3$)$_3$] is monoclinic (space group $P2_1/m$) with two formula units per primitive cell.~\cite{volkova-10, gagelmann-11} For this crystal symmetry the factor group analysis yields 45 Raman-active phonon modes: $\Gamma_{Raman} = 25 \cdot A_g + 20 \cdot B_g$. The corresponding Raman tensors are given by:
\begin{center}

\mbox{A$_{g}$=$\begin{pmatrix} a & d & 0\\ d & b & 0\\ 0 & 0 & c\\
\end{pmatrix}$

, B$_{g}$=$\begin{pmatrix} 0 & 0 & e\\ 0 & 0 & f\\ e & f & 0\\ \end{pmatrix}$.}

\end{center}
Raman scattering experiments performed in different polarizations at room temperature ($T = 295$ K) within the $ab$ plane of the crystal do not reveal a noteworthy dependence on the polarization configurations. This is probably due to multiple reflections of the incident light in the transparent crystals. Therefore we have removed the analyzer to obtain maximum scattering intensity with ($xx + yx$) polarization in all following measurements.

\begin{figure}
\centering
\includegraphics[width=8cm]{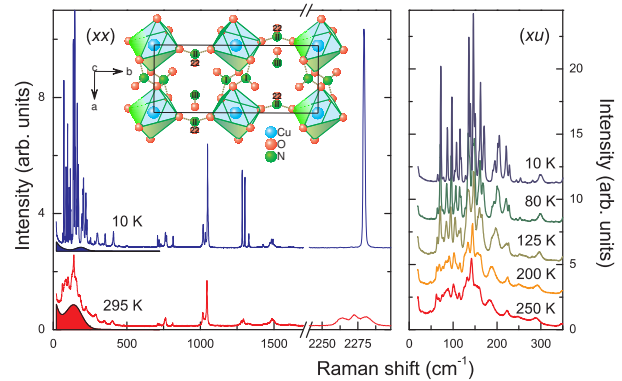}
\caption{\label{overview+structure}(Color online) Left panel: Raman spectra of (NO)[Cu(NO$_3$)$_3$] obtained in ($xx$) polarization at 10 K and 295 K. The filled background emphasizes the temperature evolution of the observed scattering continuum. The crystallographic structure of (NO)[Cu(NO$_3$)$_3$] is shown as an inset within the $ab$-plane. The N(II)O$_3^-$ nitrate groups connect Cu chains along the $b$-axis, while two N(I)O$_3^-$ nitrate groups form more complex but weaker interchain connections along the $a$-axis. Right panel: Temperature evolution of the Raman spectra in the low-frequency range in ($xu$) polarization.}
\end{figure}

Raman scattering spectra collected at 295 K and 10 K in the frequency range up to 2400 cm$^{-1}$ are presented in the left panel of Fig.~\ref{overview+structure}. According to the three structural subunits the lattice vibrations in (NO)[Cu(NO$_3$)$_3$] can be separated into distinctive groups of frequencies corresponding to external translational and rotational modes (70 -- 200 cm$^{-1}$), internal modes of CuO$_6$ (200 -- 600 cm$^{-1}$), nitrate groups NO$_3^-$ (700 -- 1500 cm$^{-1}$)~\cite{kamboures-08} as well as NO$^+$ groups~\cite{givan-89} ($\approx$2280 cm$^{-1}$). Fitting the spectrum with Lorentzian profiles allows to distinguish 41 phonon modes in the entire frequency region. This is in good agreement with the factor group analysis. It is noteworthy that there exists a well defined low energy onset of the phonon modes at E$_{low}\approx$70 cm$^{-1}$ = 100 K, which coincides with the above mentioned $T^*$ from $\chi(T)$. There is a pronounced temperature dependence of the phonons as well as of a broad scattering continuum. In the following we will study both effects in detail.

\begin{figure}
\centering
\includegraphics[width=8cm]{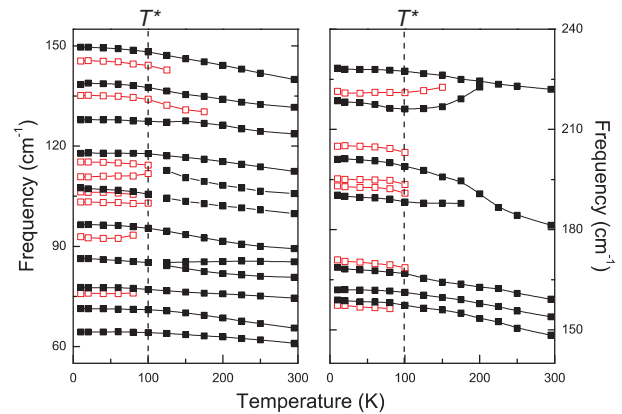}
\caption{\label{overview-frequency}(Color online) Temperature dependence of the low energy phonon frequencies. The red squares correspond to phonon modes that appear at low temperatures, $T < T^* \sim 100$ K, the latter given by a dashed vertical line. Lines are guides to the eye}
\end{figure}

In the right panel of Fig.~\ref{overview+structure} the temperature development of the spectra in the low frequency range is plotted. Strong changes as function temperature involve frequency, intensity and linewidth. In order to analyze and separate the temperature dependence of the phonon modes from the broader magnetic scattering, all spectra were analyzed by fitting them to lines with Lorentzian profiles.

In Fig.~\ref{overview-frequency} we illustrate shifts in phonon frequency in the low energy region as function of temperature. Focussing here on the modes at around 190 and 220 cm$^{-1}$ (right panel) it is obvious that there exists a gradual structural evolution. The onset of this effect is above 200 K. The strongest changes including the appearance of new modes are at $T<T^* \approx 100$ K. This is the characteristic temperature where the magnetic susceptibility $\chi(T)$ shows a broad maximum. At low temperatures the total number of phonons exceeds the expected 45 modes from the factor group analysis, i.e. we have to consider a reduced local symmetry in (NO)[Cu(NO$_3$)$_3$].

\begin{figure}
\centering
\includegraphics[width=8cm]{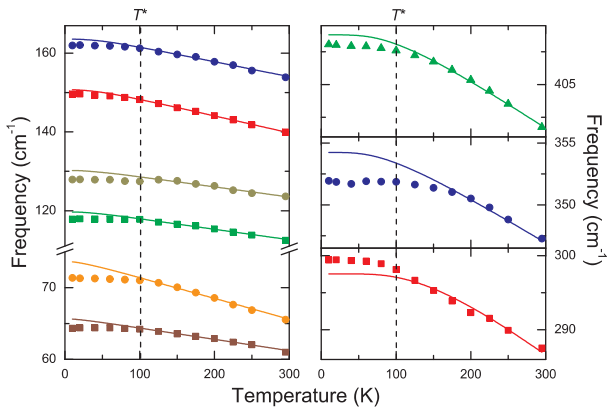}
\caption{\label{frequency}(Color online) Analysis of the phonon frequency shifts (symbols) using a model (lines) based on phonon-phonon decay processes.~\cite{balkanski-83} The corresponding modes show deviations for temperatures $T<T^*$, especially for higher frequencies, as described in the text.}
\end{figure}

Figure~\ref{frequency}, left panel, shows the temperature evolution of some external translational and rotational phonon frequencies together with a fit based on phonon-phonon decay processes,~\cite{balkanski-83}
$\omega_{Ph}(T) = \omega_0 + C \left( 1+ \frac{2}{\exp(\hbar \omega_0 / k_B T)-1}\right)$. This fit approximates the anharmonic phonon contribution. The experimental data deviate in the low temperature regime, setting in for $T \leq T^*$ and the modes have a lower frequency than expected.
Also internal modes of CuO$_6$, expected in the range 200 -- 600 cm$^{-1}$, show an anomalous temperature dependence, see right panel of Figure~\ref{frequency}. Noteworthy is that positive as well as negative deviations from the anharmonic fit exist.

\begin{figure}
\centering
\includegraphics[width=8cm]{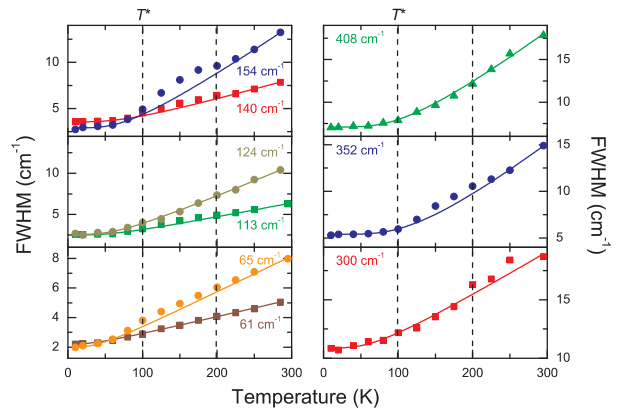}
\caption{\label{linewidth}(Color online) Analysis of the phonon linewidths (FWHM, symbols) using a phenomenological model (lines). Several modes show deviations for temperatures 200 K $< T < T^*$.}
\end{figure}

The linewidths (full width at half maximum, FWHM) of selected modes in the frequency region of external (internal) vibrations are depicted in the left (right) panel of Fig.~\ref{linewidth}. Our analysis shows that many phonons are anomalous, i.e. they show variations as a function of temperature that are not consistent with anharmonicity. The solid lines in the temperature dependent linewidths are a fit using a phenomenological model:
$\Gamma(T) = \Gamma_0 \frac{1 + d_j}{\exp(\hbar \omega_0 / k_B T)-1}$,
with $d_j$ being a mode dependent fit parameter. The phonon linewidths show a general decrease with reduced temperature. This is due to anharmonicity and phonon--phonon scattering. Noteworthy, however, is the superimposed change of curvature leading to an additional drop of linewidth for $T<T^* \approx 100$ K of the modes at 140, 154 and weaker for the one at 65 cm$^{-1}$.

\begin{figure}
\centering
\includegraphics[width=8cm]{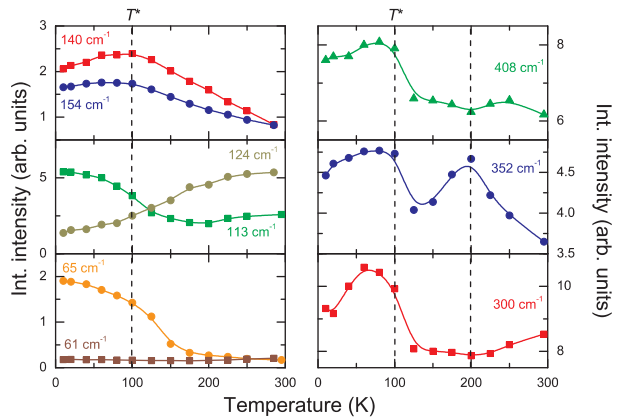}
\caption{\label{intensity}(Color online) Integrated intensity over temperature of selected phonon modes. Lines are guides to the eye.}
\end{figure}

Fig.~\ref{intensity} plots the Bose corrected integrated intensity, in analogy to frequency and linewidth of the phonon modes. It is noteworthy that the intensity evolution is different for each mode without an obvious dependence on the frequency of the mode. The observed effect is therefore not related to the scattering volume, i.e. by a change of the optical penetration depth. It has rather to be individually assigned to certain spin-phonon coupling parameters as discussed below.
The integrated intensity of the internal CuO$_6$ modes (right panel, Fig.~\ref{intensity}) shows evidence for a second characteristic temperature at 200~K marked by a dashed lines.

The integrated Raman scattering intensity of a phonon mode consists beside the scattering volume of two contributions.~\cite{sherman-03} The dipole matrix elements arise from a phonon-induced polarization of the wave function and these elements vary exponentially with the interatomic distance. The other contribution is related to displacement induced changes in the band energies. As the effects in (NO)[Cu(NO$_3$)$_3$] are specific for each phonon line they can not be explained in terms of lattice thermal contraction. Therefore, dynamic changes of ionic distances are the most likely origin for the observed variation of phonon intensity.

There exists a further Raman scattering signal in (NO)[Cu(NO$_3$)$_3$] at $T > T^*$ with a much larger linewidth than the phonon scattering, see the solid background in the left panel of Fig.~\ref{overview+structure}. This continuum shows a finite energy maximum ($E_{max} \approx 200$ cm$^{-1}$). We have no evidence for disorder that could broaden phonon lines. Therefore we attribute the continuum to magnetic Raman scattering.~\cite{lemmens-rev}



\begin{figure}
\centering
\includegraphics[width=8cm]{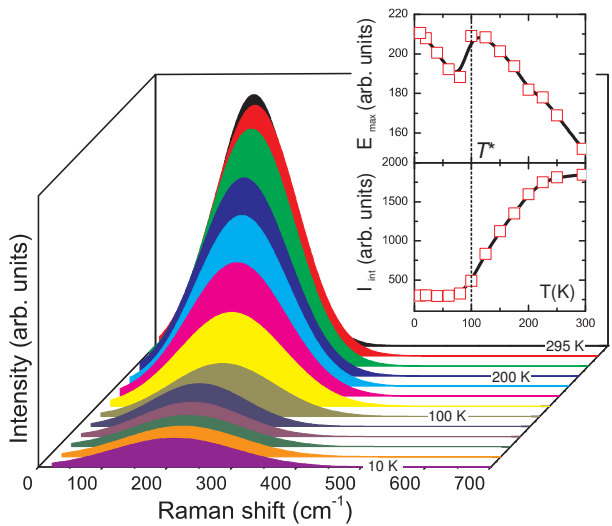}
\caption{\label{continuum}(Color online) Magnetic Raman scattering with a finite energy maximum $E_{max}(T)$ determined from fitting the data after subtracting all phonon lines. The curves were obtained at 10, 20, 40, 60, 80, 100, 125, 150, 175, 200, 225, 250, and 295 K. The upper inset shows $E_{max}(T)$ and the lower inset the integrated intensity I$_{int}(T)$.}
\end{figure}

The intensity of the continuum is strongly reduced at low temperatures, see lower inset of Fig.~\ref{continuum}. We have to admit that the continuum overlaps with many sharp phonon lines. Therefore it is difficult to determine whether its intensity is reduced exactly to zero or to a finite magnitude. The upper inset shows the energy of the maximum. There is a continuous shift by approximately 10$\%$ of its position to higher energy with decreasing temperatures. Superimposed on this shift there is a jump at $T^*$.

In a 1D $s=1/2$ spin system Raman scattering on spinon excitations may be observed as a broad maximum with a typical energy of $E_{max} = 2J$, with $J$ the exchange coupling along the chain direction. From the maximum in Fig.~\ref{continuum} we derive a coupling constant of $J_b$$\approx$150~K. This value compares very well with previous values from thermodynamics,~\cite{volkova-10} $J$=170 K, and still reasonably well with $J_b$$\approx$200~K from DFT calculations.~\cite{janson-10} Such magnetic scattering is only observed if the Raman exchange scattering Hamiltonian $\mathcal H_R = \sum s \cdot s$ does not commute with the magnetic exchange Hamiltonian of the spin systems, i.e. additional couplings causing frustration or a dimerization is mandatory.~\cite{gros03}

The strong decrease of scattering intensity for $T < T^*$ K is not expected. In contrast, spinon and two-magnon scattering is only weakly temperature dependent and only broadened by thermal fluctuations with $T$ comparable to the coupling constant. This has been shown, e.g. for magnetic Raman scattering in the kagome lattice.~\cite{wulferding-10} The temperature effect could be related to a suppression of the free spinons by opening a static or dynamic gap. In the spin Peierls system CuGeO$_3$ a similar depression has been observed with decreasing temperatures in proximity to the dimerized phase.~\cite{gros03} Although $\chi(T)$ is smaller than expected, there is no pronounced effect as expected due to the freezing out of triplet excitations. Nevertheless, the frequency shift of the continuum seams to indicate a continuous change of the exchange parameters with $T$ in the overall temperature range as well as a discontinuous, first-order like change at $T^*$.

\section{Discussion}

In the following we will discuss the implications of the phonon anomalies for the structure of (NO)[Cu(NO$_3$)$_3$]. Then possible magnetic ground states consistent with the observed magnetic Raman scattering will be evaluated.

In discussing phonon anomalies in (NO)[Cu(NO$_3$)$_3$] it is important to realize that its structural data~\cite{znamenkov-04} evidences a remarkable flexibility of the NO$_3^-$ and NO$^+$ subunit positions, see inset of Fig.~\ref{overview+structure}. While the Cu positions centered in CuO$_6$ octahedra are firmly fixed and may play a stabilizing role, other structural units are located at the general $4f$ ($x,y,z$) crystallographic positions in which all $x,y,z$ coordinates are free. This concerns all atoms of the two N(I)O$_3^-$ subunits or N(II) and O(22) atoms from the N(II)O$_3^-$ subunits at the $2e$ ($x,1/4,z$) positions in which two $x-$ and $z$-coordinates are free, and all atoms from NO$^+$ subunits. Here we use the notation from Ref.~\cite{znamenkov-04} for the labeling of oxygen and nitrogen atoms. The NO$_3^-$ complexes are therefore allowed to rotate as a whole and NO$^+$ complexes can rotate within the $ac$-plane without changing any Cu-O bond length.

A thermal population of the low frequency external translational and rotational modes of the NO$_3^-$ and NO$^+$ subunits thereby enhances the anharmonicity of all optical phonons as decay channels are opened. As these phonon modes have a low energy onset at E$_{low}\approx$ 70 cm$^{-1}$ = 100 K, most of the phonon anomalies are characterized by this temperature as well. Such a mechanism via lattice anharmonicity also takes into account that the observed frequency shifts are rather moderate, around or below the 1\% range. Therefore we suggest a tendency for local or fluctuating lattice distortions in (NO)[Cu(NO$_3$)$_3$] that could eventually lead to a crystallographic distortion with reduced site symmetry.

We have evidence for such a kind of dynamical instability also from the high frequency phonon modes. The existence of a few dynamically stable positions at high temperature is observed as a splitting of the characteristic NO$^+$ mode at around 2000 cm$^{-1}$. This splitting disappears at low temperatures (see high frequency range in the left panel of Fig.~\ref{overview+structure}). Moreover, the phonon frequency $\gamma$(NO$^+$) is located at low temperatures at 2280 cm$^{-1}$. This frequency reflects an absence (or minimum) of interaction between the NO$^+$ units and Cu atoms. In the opposite case, a significant decrease of the $\gamma$(NO$^+$) frequency should be observed as was demonstrated for several NO complexes.~\cite{lewis-58}

Taking this structural flexibility and dynamics into account we differentiate two effects on the magnetic properties of (NO)[Cu(NO$_3$)$_3$]. CuO$_6$ octahedra determine with their Cu-O bond lengths the strength and with their distortions a possible dimerization of the exchange coupling along the \emph{b}-axis. This coupling establishes the 1D magnetic correlations of this compound, which is dominating. In contrast, the NO$_3^-$ complexes mediate the direct and diagonal frustrating couplings of the spin chains with each other.~\cite{janson-10} These couplings are critical for establishing different kinds of quantum disordered ground states \cite{wang11} and whether (NO)[Cu(NO$_3$)$_3$] realizes the Nersesyan-Tsvelik model.

The phase diagram of 2D $J_1/J_2$ quantum spin systems has been investigated theoretically \cite{kalz2011} as function of frustrated couplings and thermal fluctuations in detail. Despite several open questions it is clear that quantum disordered phases, e.g. of the Nersesyan-Tsvelik model, have a rather small phase space. Such phases are enclosed by magnetically ordered or dimerized states with different low energy excitations. Thermal fluctuations destabilize long range order and thereby enlarge the phase space of the exotic phases \cite{kalz2011}. We attribute the intensity variation of the spinon continuum to a temperature induced shift of the system to the boundary of an ordered state. In this sense the behavior of (NO)[Cu(NO$_3$)$_3$] is unconventional and could be compared with a spin chain system that is coupled to phonons. In the latter system a dimerization transition (e.g. a spin-Peierls transition at $T_{SP}$) opens a spin gap and suppresses spinon excitations at low temperatures. Such a gap, however, concerns both singlet and triplet excitations and it is at a temperature much smaller than the characteristic energy scale of the spin system. In (NO)[Cu(NO$_3$)$_3$] we observe the onset of the suppression of spinon scattering at around 250~K, a temperature of approximately 2$J_b$. A modeling of (NO)[Cu(NO$_3$)$_3$] should therefore consider a 2D exchange topology including lattice fluctuations. A purely 1D spin chain model is not in agreement with the observation of a spinon continuum in Raman scattering. Additional spectroscopic and structural investigations of (NO)[Cu(NO$_3$)$_3$] also at lower temperatures, e.g. using neutron scattering, would be needed to further clarify this point.

\section{Summary}
In conclusion, we observe phonon anomalies in (NO)[Cu(NO$_3$)$_3$] promoted by the weakly coupled structural building blocks of the compound. The characteristic temperature of these anomalies and of the magnetic susceptibility coincide. In addition, a broad, temperature dependent scattering continuum of magnetic origin is observed and attributed to free spinon excitations. The position of this maximum is related to the major exchange coupling $J_b$$\approx$150~K along the chain direction in good agreement with previous investigations. We attribute further anomalies to a pronounced anharmonicity and dynamics of NO$_3^-$ and NO$^+$ subunits that modulate the coupling between the chains. Our observations do not directly contrast with the proposed Nersesyan-Tsvelik model. However, a refined description of the compound should take spin-phonon coupling and lattice dynamics into account.

\begin{acknowledgments}
We thank Prof. Viktor Eremenko for his continuing interest and support of our work. Furthermore we acknowledge important discussions with Yasir Iqbal and support from DFG, B-IGSM and the NTH School ``Contacts in Nanosystems''.
\end{acknowledgments}

\end{document}